\documentclass[twocolumn,showpacs,amsmath,amssymb,prb]{revtex4}
\usepackage{graphicx}% Include figure files
\usepackage{dcolumn}% Align table columns on decimal point
\usepackage{bm}% bold math

\begin{document}

\title{High-dielectric constant and wide band gap inverse silver oxide \\ 
phases of the ordered ternary alloys of SiO$_{2}$, GeO$_{2}$ and SnO$_{2}$}

\author{C. Sevik}
\email{sevik@fen.bilkent.edu.tr}
\author{C. Bulutay}
\email{bulutay@fen.bilkent.edu.tr}
 \affiliation{Department of Physics, Bilkent University, Bilkent,
 Ankara, 06800, Turkey}
\date{\today}

\begin{abstract}
High-dielectric constant and wide band gap oxides have important technological 
applications. The crystalline oxide polymorphs having lattice constant 
compatibility to silicon are particularly desirable. One recently reported 
candidate is the inverse silver oxide phase of SiO$_2$. First-principles 
study of this system together with its isovalent equivalents GeO$_{2}$, 
SnO$_{2}$ as well as their ternary alloys are performed. Within the framework 
of density functional theory both generalized gradient approximation and local 
density approximation (LDA) are employed to obtain their structural properties, 
elastic constants and the electronic band structures. To check the stability
of these materials, phonon dispersion curves are computed which indicate that
GeO$_{2}$ and SnO$_{2}$ have negative phonon branches whereas their ternary 
alloys Si$_{0.5}$Ge$_{0.5}$O$_{2}$, Si$_{0.5}$Sn$_{0.5}$O$_{2}$, and 
Ge$_{0.5}$Sn$_{0.5}$O$_{2}$ are all stable within LDA possessing dielectric 
constants ranging between 10 to 20. Furthermore, the lattice constant of 
Si$_{0.5}$Ge$_{0.5}$O$_{2}$ is virtually identical to the Si(100) surface.
The $GW$ band gaps of the stable materials are computed which restore the 
wide band gap values in addition to their high dielectric constants.
\end{abstract}

% 61.50.Ah 	Theory of crystal structure, crystal symmetry; calculations
%and modeling
% 62.20.Dc 	Elasticity, elastic constants
% 63.20.Dj 	Phonon states and bands, normal modes, and phonon dispersion
% 71.15.Mb 	Density functional theory, local density approximation, 
%gradient and other corrections
% 71.20.-b 	Electron density of states and band structure of crystalline 
%solids
% 71.20.Ps 	Other inorganic compounds
% 77.22.Ch 	Permittivity (dielectric function)

\maketitle

%\section{Introduction}
High-dielectric constant and wide band gap oxides are of general interest for the next-generation gate oxides for silicon-based electronics~\cite{robertson} and also as host matrices for nonvolatile flash memory applications.\cite{flash-mem} Amorphous oxides have been generally preferred as they are good glass-formers which tend to minimize the number of dangling bonds at the interface. In this respect, poly-crystalline oxides are undesirable as the grain boundaries cause higher leakage currents and possible diffusion paths for dopants.\cite{robertson} On the other hand, {\it crystalline} oxide grown epitaxially on silicon~\cite{xtal-oxide} can be favorable as it will result in high interface quality provided that it is lattice-matched to Si.

Very recently, Ouyang and Ching~\cite{ouyang} have reported a
high-density cubic polymorph of SiO$_2$ in the inverse Ag$_2$O structure, 
named by them as the i-phase, possessing both high dielectric constant, as 
in stishovite phase, and the lattice constant compatibility to Si(100) 
face which make it very attractive for electronic applications. In this computational work, we continue this search for the crystalline high-dielectric constant oxides with the i-phases of GeO$_2$ and SnO$_2$ as well as their ordered ternary alloys with SiO$_2$. This pursuit is in line with the International Technology Roadmap for Semiconductors where computational synthesis of novel high-dielectric materials is emphasized.\cite{itrs} We employ the well-established \textit{ab initio} framework based on the density functional theory within the generalized gradient approximation and local density approximation using pseudopotentials and a plane wave basis.\cite{martin-book}

\begin{figure}[ht!]
\includegraphics[width=6cm]{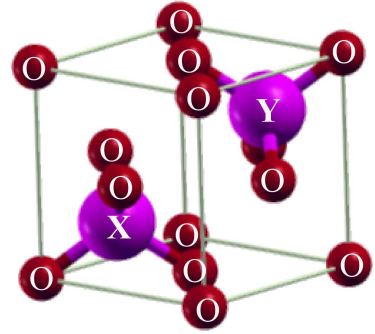}
\caption{\label{Ballstick}Ball and stick model of the i-phase ordered 
ternary alloy X$_{0.5}$Y$_{0.5}$O$_{2}$.}
\end{figure}

%\section{Computational Details}
The unit cell for the ordered ternary alloy X$_{0.5}$Y$_{0.5}$O$_{2}$ in the inverse Ag$_{2}$O structure is shown in Fig.~\ref{Ballstick}. Structural and electronic properties of the i-phase structures under consideration have been calculated within the density functional theory,\cite{martin-book} using the plane wave basis pseudopotential method as implemented in the ABINIT code.\cite{gonze} The results are obtained under the the generalized gradient approximation (GGA) and local density approximation (LDA) where for the exchange-correlation interactions we use the Teter-Pade parameterization,\cite{xcteter} which reproduces Perdew-Zunger~\cite{perdew}  (which reproduces the quantum Monte Carlo electron gas data of Ceperley and Alder\cite{ceperley}). We tested the LDA results under two different norm-conserving Troullier and Martins~\cite{tm1} type pseudopotentials, which were generated by  A. Khein and D.C. Allan (KA) and Fritz Haber Institute (FHI); for either set, the $d$ electrons were not included in the valence configuration. Our calculated values for these two types of pseudopotentials were very similar.

\begin{figure*}[t!]
\includegraphics[width=14cm]{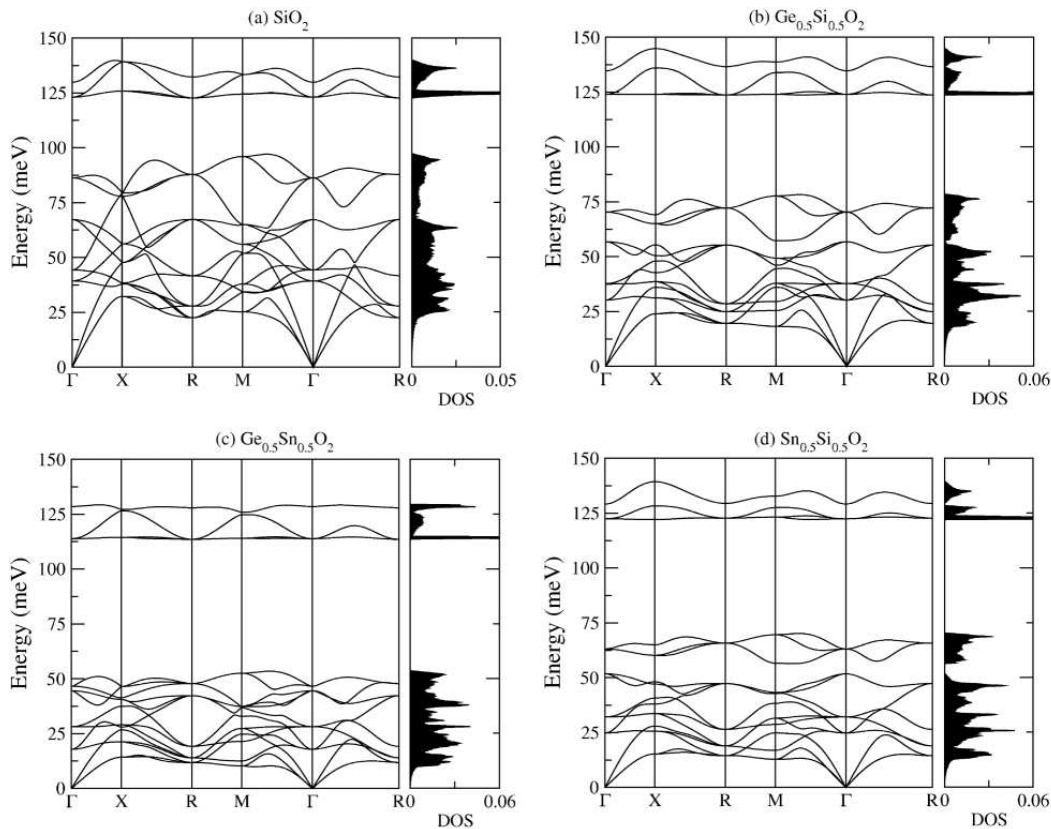}
\caption{\label{Phon}LDA phonon dispersions and the phonon DOS (a. u.) of the 
stable crystals: (a) SiO$_{2}$, (b) Ge$_{0.5}$Si$_{0.5}$O$_{2}$, 
(c) Ge$_{0.5}$Sn$_{0.5}$O$_{2}$, and (d) Si$_{0.5}$Sn$_{0.5}$O$_{2}$.}
\end{figure*}
In the course of both GGA and LDA computations, the plane wave energy cutoff and k-point sampling were chosen to assure a 0.001~eV energy convergence for all i-phase crystals. In the case of SiO$_{2}$ this demands a 65~Ha plane wave energy cutoff and 10$\times$10$\times$10 $k$-point sampling. Phonon dispersions and phonon density of states were computed by the PHON program~\cite{phon} using a $2\times 2\times 2$ supercell of 48 atoms to construct the dynamical matrix.The required forces were extracted from ABINIT. The corrected band gap values are computed by obtaining  self-energy corrections to the DFT Kohn-Sham eigenvalues in the $GW$ approximation.\cite{GW} All parameters used during the $GW$ calculation were chosen to assure a 0.001~eV energy convergence.

\begin{table}[h!]
\caption{\label{structure}First-principles LDA and GGA structural data for i-phase crystals.}
\begin{ruledtabular}
\begin{tabular}{llcccc}
Crystal&&a (\AA) &Density (gr/cm$^{3}$)&x-O (\AA)&y-O (\AA)\\
\hline
SiO$_{2}$ &LDA& 3.734&3.830&1.617& \\
&GGA& 3.801&3.633&1.646& \\
\hline
GeO$_{2}$&LDA&3.916&5.781&1.696&\\
&GGA&4.053&5.215&1.755&\\
\hline
SnO$_{2}$&LDA&4.180&6.864&1.808&\\
&GGA&4.452&5.671&1.928&\\
\hline
Ge$_{0.5}$Si$_{0.5}$O$_{2}$&LDA&3.836&4.843&1.697&1.625\\
&GGA&3.923&4.528&1.762&1.635\\
\hline
Ge$_{0.5}$Sn$_{0.5}$O$_{2}$&LDA&4.042&6.416&1.688&1.813\\
&GGA&4.250&5.522&1.748&1.932\\
\hline
Sn$_{0.5}$Si$_{0.5}$O$_{2}$&LDA&3.970&5.590&1.818&1.620\\
&GGA&4.114&5.015&1.935&1.628\\
\end{tabular}
\end{ruledtabular}
\end{table}

%\section{Results}
Using XO$_{2}$ and X$_{0.5}$Y$_{0.5}$O$_{2}$ as the generic notation, the O-X-O and O-Y-O bond angles are 109.47$^{\circ}$ and the X-O-X and X-O-Y bond angles are 180$^{\circ}$ according to the crystal construction of this cubic i-phase (cf. Fig.~\ref{Ballstick}). Other structural information such as the lattice constants and bond lengths of all i-phase crystals are listed in Table~\ref{structure}.
\begin{table}[ht!]
\caption{\label{elastic}Elastic constants and bulk modulus for each crystal.}
\begin{ruledtabular}
\begin{tabular}{llccccc}
Crystal&&$C_{11}$(GPa)&$C_{12}$(GPa)&$C_{44}$(GPa)&$B$(GPa) \\
\hline
\hline
SiO$_{2}$&LDA&383.6&260.0&243.0&301\\
&GGA&354.3&232.1&227.9&273\\
\hline
GeO$_{2}$&LDA&297.0&231.2&175.6&253\\
\hline
SnO$_{2}$&LDA&208.9&185.5&113.9&193\\
\hline
Ge$_{0.5}$Si$_{0.5}$O$_{2}$&LDA&349.4&253.2&200.0&285\\
&GGA&292.8&203.9&161.8&234\\
\hline
Ge$_{0.5}$Sn$_{0.5}$O$_{2}$&LDA&255.4&210.8&106.3&226\\
\hline
Sn$_{0.5}$Si$_{0.5}$O$_{2}$&LDA&277.5&217.4&103.9&237\\
&GGA&238.3&183.0&202.8&201\\
\end{tabular}
\end{ruledtabular}
\end{table}
\begin{figure*}[t!]
\includegraphics[width=14cm]{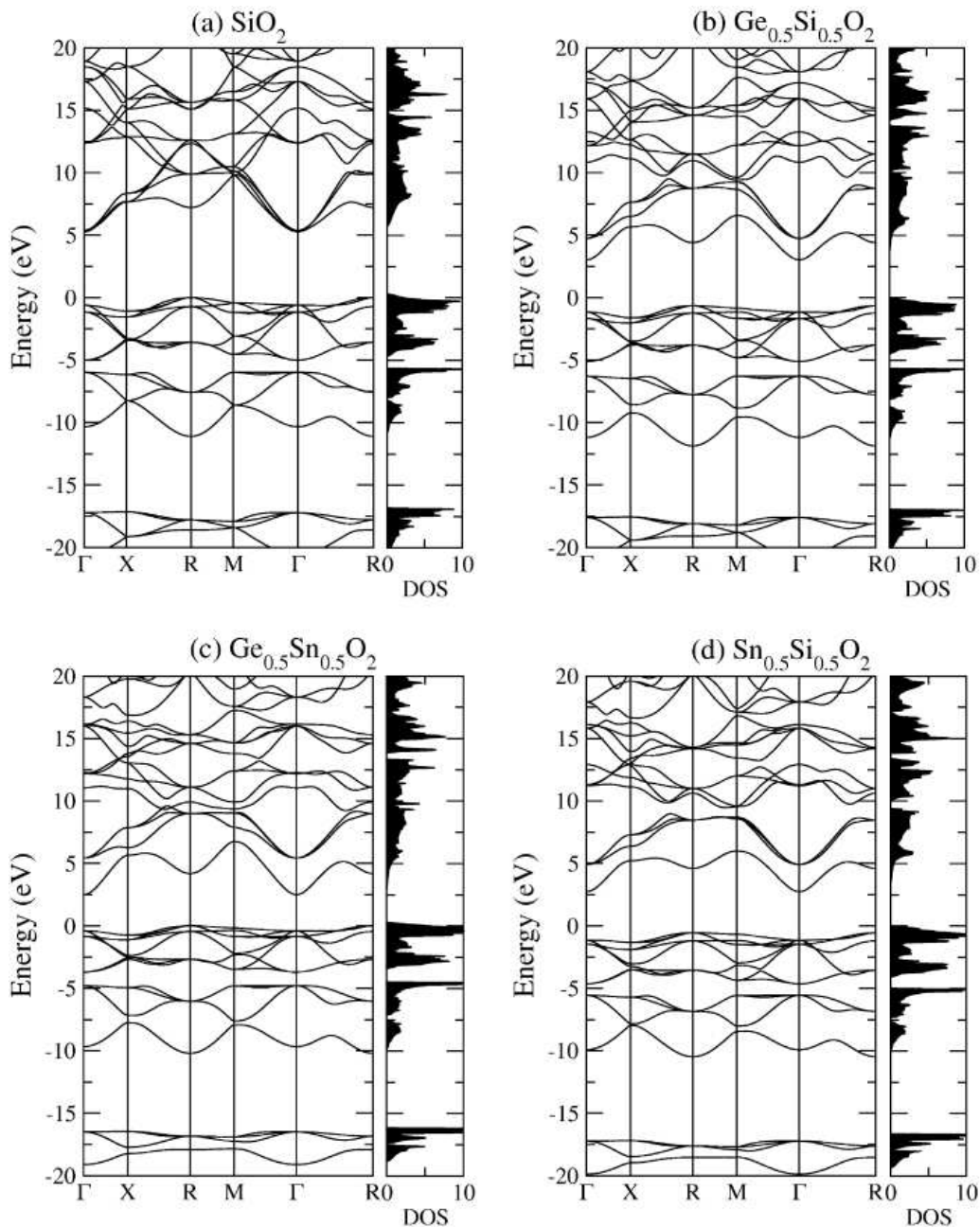}
\caption{\label{Band}LDA electronic band structure and DOS (States/eV cell) of 
i-phase (a) SiO$_{2}$, (b) Ge$_{0.5}$Si$_{0.5}$O$_{2}$, 
(c) Ge$_{0.5}$Sn$_{0.5}$O$_{2}$, and (d) Sn$_{0.5}$Si$_{0.5}$O$_{2}$.}
\end{figure*}
The Si(100) surface lattice constant is about 3.83~{\AA}, therefore according to LDA results Si$_{0.5}$Ge$_{0.5}$O$_{2}$ is of particular interest as it can be epitaxially grown on Si without any strain. According to our well-converged calculations Si$_{0.5}$Ge$_{0.5}$O$_2$ has a lower total energy compared to both SiO$_2$ and GeO$_2$, the latter itself is unstable as will be shown later; this can be taken as some indication of immunity to the phase separation of this ternary alloy into its binary compounds.

The LDA and GGA results of the three independent elastic constants and bulk modulus  for all crystals are tabulated in Table~\ref{elastic}. An important concern is the stability of these cubic phases. The requirement of mechanical stability on the elastic constants in a cubic crystal leads to the following constraints: $C_{11} > C_{12}$, $C_{11}>0$, $C_{44}>0$, and $C_{11}+2C_{12} > 0$. The elastic constants calculated by both LDA and GGA shown in Table~\ref{elastic} satisfy these stability conditions. Furthermore, we compute the LDA and GGA phonon dispersion curves of these structures using the PHON program.\cite{phon} First, to verify the validity of the results of the PHON program we compute the phonon dispersions of the SiO$_{2}$ and GeO$_{2}$ by using both PHON and ANADDB extension of the ABINIT code.\cite{gonze} There exists a good agreement between two calculations. Next, we calculate the phonon dispersions of the all i-phase crystals via PHON program with forces obtained from LDA and GGA. It is observed that SiO$_{2}$ is at least locally stable whereas GeO$_{2}$ and SnO$_{2}$ contains negative phonon branches which signal an instability of these phases. As for their alloy, Ge$_{0.5}$Sn$_{0.5}$O$_{2}$, according to LDA this material is stable whereas within GGA it comes out as unstable. For the stable structures the LDA phonon dispersions and the associated phonon density of states (DOS) are shown in Fig.~\ref{Phon}.

\begin{table}[h!]
\caption{\label{dielectric}LDA and GGA dielectric permittivity tensor for the stable crystals.}
\begin{ruledtabular}
\begin{tabular}{llcc}
Crystal&&$\epsilon_{xx}^{0}=\epsilon_{yy}^{0}$=$\epsilon_{zz}^{0}$&
$\epsilon_{xx}^{\infty}=\epsilon_{yy}^{\infty}$=$\epsilon_{zz}^{\infty}$\\
\hline
SiO$_{2}$&LDA&9.857&3.285\\ 
&GGA&9.970&3.303\\ \hline
Ge$_{0.5}$Si$_{0.5}$O$_{2}$&LDA&11.730&3.416 \\ 
&GGA&14.383&3.585 \\ \hline
Ge$_{0.5}$Sn$_{0.5}$O$_{2}$&LDA&19.415&3.527 \\ \hline
Sn$_{0.5}$Si$_{0.5}$O$_{2}$&LDA&12.883&3.360 \\ 
&GGA&18.096&3.711 \\ 
\end{tabular}
\end{ruledtabular}
\end{table}

For the stable systems, the static and high-frequency dielectric constants are
listed in Table~\ref{dielectric}. The static dielectric constants falling in 
the range between 10 to 20 suggest that these are moderately high dielectric 
constant crystals. It can be observed that GGA yields systematically higher 
values for the dielectric constants of these structures.
Employing KA pseudopotentials, 
the LDA band structure for the crystals are displayed 
along the high-symmetry lines in Fig.~\ref{Band} including the electronic DOS.
The widths of the valence bands get progressively narrowed from Fig.~\ref{Band}(a) 
to (d), i.e., from SiO$_2$ to Sn$_{0.5}$Si$_{0.5}$O$_2$. 
For all of the i-phase crystals under consideration including the unstable 
ones the conduction band 
minima occur at the $\Gamma$ point whereas the valence band maxima are 
located at $R$ point making them indirect band gap semiconductors. 
As tabulated in Table~\ref{egap}, the direct band gap values are only
marginally above the indirect band gap values. Again GGA systematically 
yields narrower band gaps compared to LDA.

A renown artifact of LDA is that for semiconductors and 
insulators band gaps are underestimated.\cite{martin-book} In this work, 
the corrected band gap values are also provided by $GW$ approximation.
As there are different $GW$ implementations we briefly highlight the particular
methodology followed in the ABINIT code. First, a converged ground 
state calculation (at fixed lattice parameters and atomic positions) is done 
to get self-consistent density and potential, and Kohn-Sham eigenvalues and 
eigenfunctions at the relevant band extrema $k$-points as well as on a 
regular grid of $k$-points. Next, on the basis of these available Kohn-Sham 
data, the independent-particle susceptibility matrix $\chi_0$ is computed on 
a regular grid of $q$-points, for at least two frequencies (usually, 
zero frequency and a large pure imaginary frequency - on the order of the 
plasmon frequency, a dozen of eV). Finally, the Random Phase Approximation 
susceptibility matrix, $\chi$, the dielectric matrix $\epsilon$ and its 
inverse $\epsilon^{-1}$ are computed. On this basis, the self-energy, $\Sigma$ 
matrix element at the given $k$-point is computed to derive the $GW$ eigenvalues 
for the target states at this $k$-point. Note that this $GW$ correction is achieved 
as a one-shot calculation (i.e., no overall self-consistency) hence, our results 
technically corresponds to $G_0W_0$ which has been the standard 
approach as originally proposal by Hedin.\cite{hedin} The $GW$ correction as can be observed 
from Table~\ref{egap} restores the wide band gap values; this feature is essential 
for these materials to provide sufficient confinement to carriers of the narrow band gap 
semiconductors such as silicon.

\begin{table}[ht!]
\caption{\label{egap}Indirect ($E_{g}$) and direct ($E_{g, \Gamma}$) band
gaps for each i-phase crystal within LDA, GGA, and for the stable structures 
the $GW$ approximation (GWA).}
\begin{ruledtabular}
\begin{tabular}{llcc}
Crystal&&$E_{g}$ (eV)&$E_{g, \Gamma}$ (eV)\\\hline
SiO$_{2}$&LDA&5.269&5.870\\
&GGA&4.584&5.155\\
&GWA&7.283&7.964\\\hline
GeO$_{2}$&LDA&2.402&2.511\\\hline
SnO$_{2}$&LDA&2.285&2.670\\\hline
Ge$_{0.5}$Si$_{0.5}$O$_{2}$&LDA&3.666&4.179\\
&GGA&2.558&3.005\\
&GWA&5.943&6.513\\\hline
Ge$_{0.5}$Sn$_{0.5}$O$_{2}$&LDA&2.487&2.900\\
&GGA&0.767&0.865\\
&GWA&4.533&4.972\\\hline
Sn$_{0.5}$Si$_{0.5}$O$_{2}$&LDA&3.292&3.900\\
&GGA&1.763&2.304\\
&GWA&5.484&6.153\\
\end{tabular}
\end{ruledtabular}
\end{table}

%\section{Conclusions}
We have also considered the i-phase of PbO$_2$ which turned out to be 
unstable and hence its \textit{ab initio} data are not included. 
In this work, we do not consider the thermodynamic stability of these i-phase oxides. 
However, for 
technological applications rather than bulk systems the epitaxial growth 
conditions become more critical.\cite{hubbard} A promising direction can be 
the finite temperature investigation~\cite{pasquarello} of these i-phase 
isovalent structures on Si(100) surfaces using large number of monolayers.

This first-principles study suggests that the i-phases of GeO$_{2}$ and 
SnO$_{2}$ are unstable whereas SiO$_2$, Si$_{0.5}$Ge$_{0.5}$O$_{2}$, 
Si$_{0.5}$Sn$_{0.5}$O$_{2}$ are particularly promising due to their 
high dielectric constants as well as wide band gaps as 
restored by the $GW$ correction. Moreover, they are lattice-matched 
to Si(100) face, especially for the case of Si$_{0.5}$Ge$_{0.5}$O$_{2}$. 
We believe that these findings can further boost the research on the 
crystalline oxides.

%\begin{acknowledgments}
This work has been supported by the European FP6 Project SEMINANO with
the contract number NMP4 CT2004 505285. We would like to thank O. G\"ulseren, 
R. Eryi{\u g}it, T. G{\"u}rel, D. {\c C}ak{\i}r and T. Y{\i}ld{\i}r{\i}m 
for their useful advices. The computations were performed in part at the
ULAKB{\.I}M High Performance Computing Center.
%\end{acknowledgments}

\end{document}